\documentclass[showpacs,twocolumn,prb,aps]{revtex4}

\usepackage{graphicx}
\usepackage{dcolumn}
\usepackage{amsmath}
\usepackage{subfigure,hyperref}
\usepackage[normalem,normalbf]{ulem}

\newcommand{\be}{\begin{equation}}
\newcommand{\ee}{\end{equation}}         
\newcommand{\etal}{{\it et al. }}

\newcommand\cg[1]{ \boldsymbol{ \sf #1}}
\DeclareRobustCommand\openone{\leavevmode\hbox{\small1\normalsize\kern-.33em1}}
\newcommand\ie{{\it i.e. }}

\def\nbC{{\mathchoice {\setbox0=\hbox{$\displaystyle\rm C$}%
\hbox{\hbox to0pt{\kern0.4\wd0\vrule height0.9\ht0\hss}\box0}} 
{\setbox0=\hbox{$\textstyle\rm
C$}\hbox{\hbox to0pt{\kern0.4\wd0\vrule height0.9\ht0\hss}\box0}} 
{\setbox0=\hbox{$\scriptstyle\rm
C$}\hbox{\hbox to0pt{\kern0.4\wd0\vrule height0.9\ht0\hss}\box0}}
{\setbox0=\hbox{$\scriptscriptstyle\rm C$}\hbox{\hbox to0pt{\kern0.4\wd0\vrule
height0.9\ht0\hss}\box0}}}}
\def\nbQ{{\mathchoice {\setbox0=\hbox{$\displaystyle\rm 
Q$}\hbox{\raise 0.15\ht0\hbox
to0pt{\kern0.4\wd0\vrule height0.8\ht0\hss}\box0}} 
{\setbox0=\hbox{$\textstyle\rm Q$}\hbox{\raise
0.15\ht0\hbox to0pt{\kern0.4\wd0\vrule height0.8\ht0\hss}\box0}} 
{\setbox0=\hbox{$\scriptstyle\rm
Q$}\hbox{\raise 0.15\ht0\hbox to0pt{\kern0.4\wd0\vrule 
height0.7\ht0\hss}\box0}}
{\setbox0=\hbox{$\scriptscriptstyle\rm Q$}\hbox{\raise 0.15\ht0\hbox 
to0pt{\kern0.4\wd0\vrule
height0.7\ht0\hss}\box0}}}}
\def\nbT{{\mathchoice {\setbox0=\hbox{$\displaystyle\rm 
T$}\hbox{\hbox to0pt{\kern0.3\wd0\vrule
height0.9\ht0\hss}\box0}} {\setbox0=\hbox{$\textstyle\rm 
T$}\hbox{\hbox to0pt{\kern0.3\wd0\vrule
height0.9\ht0\hss}\box0}} {\setbox0=\hbox{$\scriptstyle\rm 
T$}\hbox{\hbox to0pt{\kern0.3\wd0\vrule
height0.9\ht0\hss}\box0}} {\setbox0=\hbox{$\scriptscriptstyle\rm T$}\hbox{\hbox
to0pt{\kern0.3\wd0\vrule height0.9\ht0\hss}\box0}}}}
\def\nbS{{\mathchoice {\setbox0=\hbox{$\displaystyle     \rm 
S$}\hbox{\raise0.5\ht0%
\hbox to0pt{\kern0.35\wd0\vrule height0.45\ht0\hss}\hbox 
to0pt{\kern0.55\wd0\vrule
height0.5\ht0\hss}\box0}} {\setbox0=\hbox{$\textstyle        \rm 
S$}\hbox{\raise0.5\ht0%
\hbox to0pt{\kern0.35\wd0\vrule height0.45\ht0\hss}\hbox 
to0pt{\kern0.55\wd0\vrule
height0.5\ht0\hss}\box0}} {\setbox0=\hbox{$\scriptstyle      \rm 
S$}\hbox{\raise0.5\ht0%
\hboxto0pt{\kern0.35\wd0\vrule height0.45\ht0\hss}\raise0.05\ht0%
\hbox to0pt{\kern0.5\wd0\vrule height0.45\ht0\hss}\box0}} 
{\setbox0=\hbox{$\scriptscriptstyle\rm
S$}\hbox{\raise0.5\ht0%
\hboxto0pt{\kern0.4\wd0\vrule height0.45\ht0\hss}\raise0.05\ht0%
\hbox to0pt{\kern0.55\wd0\vrule height0.45\ht0\hss}\box0}}}}
\def\nbZ{{\mathchoice {\hbox{$\sf\textstyle Z\kern-0.4em Z$}} 
{\hbox{$\sf\textstyle Z\kern-0.4em Z$}}
{\hbox{$\sf\scriptstyle Z\kern-0.3em Z$}} 
{\hbox{$\sf\scriptscriptstyle Z\kern-0.2em Z$}}}}
\usepackage{graphicx}
\usepackage{dcolumn}
\usepackage{amsmath}

\begin{document}

\title[Short Title]{XY frustrated systems: continuous exponents in
discontinuous phase transitions.} 

\author{M. Tissier} \email{tissier@lpthe.jussieu.fr}
\affiliation{Laboratoire de Physique Th\'eorique et Mod\`eles Statistiques, 
Universit\'e Paris Sud, Bat. 100, 91405 Orsay Cedex, France.}%
\author{B. Delamotte} \email{delamotte@lpthe.jussieu.fr}
\affiliation{Laboratoire de Physique Th\'eorique et Hautes
Energies, Universit\'es Paris VI-Pierre et Marie Curie - Paris
VII-Denis Diderot, 2 Place Jussieu, 75252 Paris Cedex 05, France.  }%
\author{D. Mouhanna} \email{mouhanna@lpthe.jussieu.fr}
\affiliation{Laboratoire de Physique Th\'eorique et Hautes
Energies, Universit\'es Paris VI-Pierre et Marie Curie - Paris
VII-Denis Diderot, 2 Place Jussieu, 75252 Paris Cedex 05, France.  }%
\date{\today}

\begin{abstract}

XY frustrated magnets exhibit an unsual critical behavior: they
display scaling laws accompanied by {\it nonuniversal} critical exponents
and a {\it negative} anomalous dimension. This suggests that they
undergo weak first order phase transitions. We show that all 
perturbative  approaches that have been used to
investigate XY frustrated magnets fail to reproduce these
features. Using a nonperturbative approach based on the concept of
effective average action, we are able to account for this {\it
nonuniversal scaling} and to describe qualitatively and, to some
extent, quantitatively the physics of these systems.

\end{abstract}

\pacs{75.10.Hk,11.10.Hi,11.15.Tk,64.60.-i}

\maketitle

\section{Introduction}

After twenty-five years of intense activity, the physics of XY and
Heisenberg frustrated systems is still the subject of a great
controversy concerning, in particular, the nature of their phase
transitions in three dimensions (see for instance Ref.\,\onlinecite{diep94} for a review). On the one hand, a recent
high-order perturbative calculation~\cite{pelissetto00,pelissetto01a}
predicts in both cases a stable fixed point in three dimensions and,
thus, a second order phase transition. On the other hand, a
nonperturbative approach, the effective average action method, based
on a Wilson-like Exact Renormalization Group (ERG) equation, leads to
first order transitions~\cite{tissier00}. Actually, it turns out that,
in the Heisenberg case, these two theoretical approaches are almost
equivalent from the experimental viewpoint (see however Ref.\,\onlinecite{tissier02b}). Indeed, within the ERG approach, the
transitions are found to be {\it weakly} of first order and
characterized by very large correlation lengths and pseudo-scaling
associated with pseudo-critical exponents close to the exponents
obtained within the perturbative approach. This occurence of
pseudo-scaling and quasi-universality has been explained within ERG
approaches by the presence a local minimum in the speed of the
flow~\cite{zumbach94,tissier00}, related to the presence of a complex
fixed point with small imaginary parts, called pseudo-fixed
point\cite{zumbach94}.

XY frustrated magnets are rather different from this point of view
since their nonperturbative RG flows display neither a fixed point nor
a minimum. We show in this article that they nevertheless {\it
generically} exhibit large correlation lengths at the transition and
thus, pseudo-scaling, but now {\it without} quasi-universality. More
precisely, we show that quantities like correlation length and
magnetization behave as powers of the reduced temperature on several
decades. A central aspect of our approach is that, although the RG
flow displays neither a fixed point nor a minimum, it remains
sufficiently slow in a large domain in coupling constant space to
produce {\it generically } large correlation lengths and scaling
behaviors. We argue that our approach allows to account for the
striking properties of the XY frustrated magnets like the XY Stacked
Triangular Antiferromagnets (STA) such as CsMnBr$_3$, CsNiCl$_3$,
CsMnI$_3$, CsCuCl$_3$, as well as XY helimagnets such as Ho, Dy and
Tb, which display scaling at the transition {\it without} any evidence
of universality. Our conclusions are in marked contrast with those
drawn from the perturbative approach of Pelissetto {\etal}~\cite{pelissetto00,pelissetto01a} which leads to predict a second
order phase transition for XY frustrated magnets.

\section{The STA model and its long distance effective hamiltonian}

\begin{figure}[htbp] 
\begin{center}
\includegraphics[width=.8\linewidth,origin=tl]{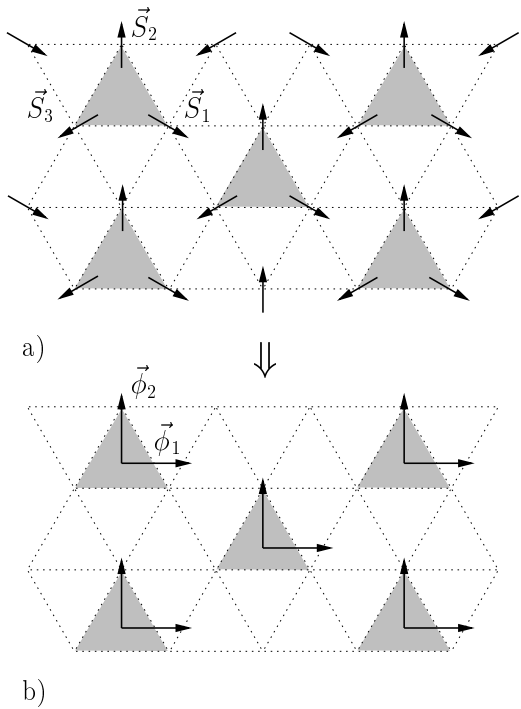}\hfill%
\end{center}
\caption{The ground state configurations a) of the spins on the
triangular lattice and b) of the order parameter made of two
orthonormal vectors. The three-dimensional structure of the ground
state is obtained by piling these planar configurations.}
\label{triangulaire}
\end{figure}

The prototype of XY frustrated systems is given by the STA model. It
consists of spins located on the sites of stacked planar triangular
lattices. Its hamiltonian reads:
\begin{equation}
H=\sum_{\langle ij\rangle}J_{ij}\vec{S}_i. \vec{S}_j
\label{hmicroscopique}
\end{equation}
where the $\vec{S}_i$ are two-component vectors and the sum runs on
all pairs of nearest neighbors.  The spins interact
antiferromagnetically inside the planes and either ferromagnetically
or antiferromagnetically between planes, the nature of this last
interaction being irrelevant to the long distance physics. Due to the
intra-plane antiferromagnetic interactions the system is geometrically
frustrated and the spins exhibit a 120$^\circ$ structure in the ground
state (see FIG. \ref{triangulaire}.a).
As $H$ is invariant under rotation, other ground states can be built
by rotating simultaneously all the spins.

Let us describe the symmetry breaking scheme of the STA model in the
continuum limit. In the high-temperature phase, the hamiltonian
(\ref{hmicroscopique}) is invariant under the $SO(2)\times {\nbZ}_2$
group acting in the spin space and the $O(2)$ group associated to the
symmetries of the triangular lattice{\footnote{As usual, we consider
the extension to the continuum of the discrete symmetry group of the
lattice.}}. In the low-temperature phase, the residual symmetries are
given by the group $O(2)_{\hbox{diag}}$ which is a combination of the
group acting in spin space and of the lattice group. The symmetry
breaking scheme is given by\cite{yosefin85,kawamura88}:
\begin{equation}
G=O(2)\times SO(2)\times {\nbZ}_2 \to H=O(2)_{\hbox{diag}}
\end{equation} 
and thus consists in a fully broken $SO(2)\times \nbZ_2$ group. The
${\nbZ}_2$ degrees of freedom are known as chirality variables.

Due to the 120$^\circ$ structure, the local magnetization, defined on
each elementary plaquette as:
\begin{equation}
\vec{\Sigma}=\vec{S}_1+\vec{S}_2+\vec{S}_3
\label{constraint}
\end{equation} 
vanishes in the ground state and cannot constitute the order
parameter. In fact, as in the case of colinear antiferromagnets, one
has to build the analogue of a staggered magnetization.  It is given
by a pair of two-component vectors $\vec{\phi}_1$ and $\vec{\phi}_2$
--- defined at the center $\cg x$ of each elementary cell of the
triangular lattice --- that are orthonormal in the ground
state~\cite{garel76,yosefin85,kawamura88} (see
FIG. \ref{triangulaire}.b). They can be conveniently gathered into a
square matrix:
\begin{equation}
\Phi(\cg x)=(\vec{\phi}_1(\cg
x),\vec{\phi}_2(\cg x))\ .
\label{matriceparametreordre}
\end{equation} 
Once the model is formulated in terms of the order parameter, the
interaction, originally antiferromagnetic, becomes ferromagnetic. It
is thus trivial to derive the effective low-energy hamiltonian
relevant to the study of the critical physics which writes:
\begin{equation}
H=-{J\over 2}\int d^d\cg x \, \hbox{Tr}\left(\partial ^{\;t}\!\Phi(\cg x).
\partial\Phi(\cg x)\right)
\label{continuhamil}
\end{equation} 
where $^{\;t}\!\Phi$ denotes the transpose of $\Phi$.

It is convenient to consider, in the following, a generalization of
the models (\ref{hmicroscopique}) and  (\ref{continuhamil}) to
$N$-component spins. The order parameter consists in this case in a
$N\times 2$ matrix and the symmetry-breaking scheme is thus given by
$O(N)\times O(2) \to O(N-2)\times O(2)_{\hbox{diag}}$. Frustrated
magnets thus correspond to a symmetry breaking scheme isomorphic to
$O(N)\to O(N-2)$ that radically differs from that of the usual
vectorial model which is $O(N)\to O(N-1)$. The matrix nature of the
order parameter together with the symmetry breaking scheme led
naturally in the 70's to the hypothesis of a new universality class
~\cite{garel76,yosefin85,kawamura88} ---
the ``chiral'' universality class --- gathering all materials supposed
to be described by the hamiltonian (\ref{hmicroscopique}): STA and
helimagnets. As we now show, examining the current state of
the experimental and numerical data, there is, in fact, no clear
indication of universality in the critical behavior of XY frustrated
magnets.

\section{The experimental and numerical context}
\subsection{ The experimental situation} 

 Two kinds of materials are supposed to undergo a phase transition
corresponding to the symmetry breaking scheme described above: the
STA --- CsMnBr$_3$, CsNiCl$_3$, CsMnI$_3$, CsCuCl$_3$ --- (see
Ref.\,\onlinecite{perez98} for RbMnBr$_3$) and the helimagnets: Ho, Dy
and Tb. The corresponding critical exponents are given in
Table~\ref{table_exp}.

Note first that, concerning all these data, only one error bar is
quoted in the literature, which merges systematic and statistical
errors. We start by making the hypothesis that these error bars have a
purely statistical origin. Under this assumption, we have computed the
--- weighted --- average values of the exponents and their error
bars. This is the meaning of the numbers we give in the
following. This hypothesis is however too  na\"\i ve, and we have checked
that, if we attribute a large part of the error bars quoted in
Table~\ref{table_exp} to systematic bias --- typically 0.1 for $\beta$
and 0.2 for $\nu$ ---, our conclusions still hold. We also make the
standard assumptions that the measured exponents govern the leading
scaling behavior, {\it i.e.} the determination of the critical
exponents is not significantly affected by corrections to
scaling. This is generically assumed in magnetic materials where
corrections to scaling are never needed to reproduce the theoretical
results in the range of reduced temperature reachable in experiments
\footnote{In the ferromagnetic Ising case, for instance, it is
probable that neglecting these corrections to scaling does not affect
much the determination of the exponents but is a possible source of
underestimations of the error bars when they are announced to be of
order of 1\% or less (see Ref.\,\cite{blote95}).}. This is different for
fluids where the scaling domain can be very large. Moreover, since the
error bars in frustrated systems are much larger than in the usual
ferromagnetic systems --- by a factor five to ten, see
Table~\ref{table_exp} --- neglecting corrections to scaling should not
bias significantly our analysis.
\begin{table}
\begin{tabular}{|l||r|}
\hline
CsMnBr$_3$& \hfill
$\alpha$=0.39(9), 0.40(5), 0.44(5)\\
&\hfill
$\beta$=0.21(1), 0.21(2), 0.22(2), 0.24(2), 0.25(1)\\
&\hfill
$\gamma$=1.01(8), 1.10(5); 
$\nu$=0.54(3), 0.57(3)  \\
\hline
CsNiCl$_3$&
\hfill
$\alpha$=0.342(5), 0.37(6), 0.37(8); 
$\beta$=0.243(5) \\
\hline
CsMnI$_3$ &
\hfill
$\alpha$=0.34(6)\\
\hline
CsCuCl$_3$&\hfill
$\alpha$=0.35(5);
 $\beta$=0.23(2), 0.24(2), 0.25(2)\\
\hline
Tb&
\hfill
$\alpha$=0.20(3); 
$\beta$=0.21(2), 0.23(4);
$\nu$=0.53\\
\hline
Ho &
\hfill
$\beta$=0.30(10), 0.37(10), 0.39(3), \\
&\hfill
0.39(2), 0.39(4), 0.39(4), 0.41(4)\\ 
&\hfill
$\gamma$=1.14(10), 1.24(15);
$\nu$=0.54(4), 0.57(4)\\
\hline
Dy &
\hfill
$\beta$=0.38(2), 0.39(1), 0.39$^{+0.04}_{-0.02}$\\
&\hfill
$\gamma$=1.05(7);
$\nu$=0.57(5)\\
\hline
STA &
\hfill
$\alpha$= 0.34(6), 0.43(10), 0.46(10)\\
Monte Carlo& $\beta$= 0.24(2), 0.253(10); $\gamma$=1.03(4), 1.13(5)\\
&\hfill 
 $\nu$=0.48(2), 0.50(1), 0.54(2)\\
\hline
six-loop &
\hfill
$\alpha$=0.29(9); 
$\beta$=0.31(2); 
$\gamma$=1.10(4); 
$\nu$=0.57(3)\\
\hline
\end{tabular}
\caption{Critical exponents of the XY frustrated models, from
Refs.\,2,11-15  and references therein. For CsCuCl$_3$ the transition
has been found of first order and the exponents mentioned here hold
only for a reduced temperature larger that 5.10$^{-3}$ (see
Ref.\,16).}
\label{table_exp}
\end{table}

Under these assumptions we can analyze the data. We find that there
are three striking facts:

\noindent {\it i}) {\it there are two groups of incompatible
exponents}.  The average value of $\beta$, the best measured exponent,
for CsMnBr$_3$, CsNiCl$_3$ and Tb --- called group 1 --- is given by
$\beta\sim 0.23$. It is incompatible with that of Ho and Dy ---
group~2 --- which is $\beta\sim 0.39$ (see Table~\ref{table_exp} for
details).  Note that for CsCuCl$_3$, whose exponents are compatible
with those of group 1, the transition has been found to be very weakly
of first order~\cite{weber96}.

\noindent {\it ii}) {\it the exponents vary much from compound to
compound in group 1.}  For instance, the values of $\alpha$ for
CsNiCl$_3$ and CsMnBr$_3$ are only marginally compatible.

\noindent {\it iii}) {\it the anomalous dimension $\eta$ is
significantly negative for group 1}.  For CsMnBr$_3$, the value of
$\eta$ determined by the scaling relation $\eta=2\beta/ \nu -1$ with
$\beta=0.227(6)$ and $\nu=0.555(21)$ is
$\eta=-0.182(38)$. The inclusion of the data coming from CsNiCl$_3$
and Tb does not change qualitatively this conclusion.

Several conclusions follow from the analysis of the data. From point
{\it i}), it appears that materials that are supposed to
belong to the same universality class differ as for their critical
behavior. There are essentially three ways to explain this. In the
first one, the two sets of exponents correspond to two true second
order phase transitions, each one being described by a fixed point. In
the second, one set corresponds to a true second order transition and
the other to pseudo-critical exponents associated to weakly first
order transitions. In the third, all transitions are weakly of first
order.

The first scenario can be ruled out since  $\eta$ is negative for 
group 1 (point {\it iii}))  while it cannot be so in a second order phase transition when the
underlying field theory is a Ginzburg-Landau $\varphi^4$-like theory
\cite{zinn_eta_pos}, as it is the case here \cite{yosefin85}.  The
transition undergone by CsMnBr$_3$, CsNiCl$_3$ and Tb is therefore
very likely {\it not} continuous but weakly of first order.  This
 would  explain the lack of universality for the exponents of group 1
  (point {\it ii})).

 In the second scenario, the materials of group 2 undergo a second
order phase transition --- $\eta$ is found positive there --- while
those of group 1 all undergo weakly first order transitions with
pseudo-scaling and pseudo-critical exponents. Note that although this
scenario cannot be excluded, it is quite unnatural in terms of the
usual picture of a second order phase transition. Indeed, it would
imply a fine-tuning of the microscopic coupling constants --- {\it
i.e.} of the initial conditions of the flow --- for the materials of
group 1 in such a way that they lie out of, but very close to, the
border of the basin of attraction of the fixed point governing the
critical behavior of materials of group 2.

The third scenario, that of generic weak first order behaviors for the
two groups of materials, seems even more unnatural, at least in the
usual explanation of weak first order phase transitions.

Actually, we shall provide arguments in favor of this last
scenario. Also, as we shall see in the framework of the effective
average action, the generic character of pseudo-scaling in this
scenario has a natural explanation, not relying on the concept of
fixed point. Then, no fine-tuning of parameters is required to explain
generically weak first order behaviors in frustrated systems.

\subsection{The numerical situation}
There are no convincing numerical data concerning helimagnets. For the
STA system, three different versions have been simulated:

1) the STA itself \cite{kawamura92,plumer94,boubcheur96}.

2) The STAR (Ref.\,\onlinecite{loison98}) --- with R for rigid --- which consists in
a STA where the fluctuations of the spins around their ground state
120$^\circ$ structure have been frozen. This is realized by imposing
the rigidity constraint $\vec{\Sigma}=\vec{S}_1+\vec{S}_2+\vec{S}_3=0$
at all temperatures.

3) A discretized version of the hamiltonian (\ref{continuhamil}),
called the Stiefel $V_{2,2}$ model \cite{loison98}. There, one
considers a system of dihedrals interacting ferromagnetically, which
is represented on FIG. {\ref{triangulaire}.b}.

At this stage, we emphasize that the rigidity constraint
$\vec\Sigma=0$ which is imposed in the STAR, as well as the formal
manipulations leading to the Stiefel $V_{2,2}$ model affects only the
{\it massive} --- non critical --- modes. Thus, all the STA, STAR and
Stiefel models have the same {\it critical} modes, the same symmetries
and thus the same order parameter.  One thus could expect {\it a
priori} that they all exhibit the same critical behavior.

For the STA system, scaling laws are
found\cite{kawamura92,plumer94,boubcheur96} so that a second order
behavior could be inferred. The STAR and $V_{2,2}$ models both undergo
first order transitions\cite{loison98}.  Therefore, by changing
microscopic details to go from the STA model to the STAR or $V_{2,2}$
models, the nature of transitions appears to change drastically. This
situation indicates that if STA undergoes a genuine second order phase
transition, the critical behavior of frustrated magnets in general is
characterized by a low degree of universality, a conclusion already
drawn from the experimental situation.

With these  behaviors one is brought back to the two last
scenarios proposed in the previous section: {\it i)} the behavior of
the STA system is controlled by a fixed point while the STAR and
Stiefel models lie outside its basin of attraction {\it ii)} all
systems undergo first order phase transitions.

 In fact, as shown in Ref.\,\onlinecite{loison98}, using the two
scaling relations $\eta=2\beta/ \nu -1$ and $\eta=2 -\gamma/\nu$,
$\eta$ is found to be negative in STA systems --- although less
significantly than in experiments --- for all simulations where these
calculations can be performed. One can thus suspect a (weak) first
order behavior even for the STA system. This hypothesis is
strengthened by a recent work of Itakura who has employed Monte Carlo
RG techniques in order to investigate the critical behavior of both
the STA system and the Stiefel model~\cite{itakura03}.  Using systems
with lattice sizes up to $126\times144\times126,$ he has provided
evidences for weak first order behaviors.

Let us draw first conclusions from the experimental and numerical
situations. It appears that the critical physics of frustrated magnets
cannot be explained in terms of a single --- universal --- second
order phase transition.  A careful analysis of the experimental and
numerical data seems to indicate that a whole class of materials
undergo (weak) first order phase transitions. At this stage, no
conclusion can be drawn about the existence or absence of a true fixed
point controlling the physics of some realizations of frustrated
magnets. To clarify this issue, we now present the theoretical
situation.

\section{The theoretical situation}
The early RG studies of the STA and helimagnets --- and its
generalization to $N$-component spins --- was performed in a double
expansion in coupling constant and in $\epsilon=4-d$ on the
Ginzburg-Landau-Wilson (GLW) version of the model in Refs.\,\onlinecite{jones76,bak76,garel76,yosefin85}. It has appeared that,
for a given dimension $d$, there exists a critical number of spin
component, called $N_c(d)$, above which the transition is of second
order and below which it is of first order. Naturally, a great
theoretical challenge in the study of frustrated magnets, has been the
determination of $N_c(d)$. Its value has been determined within
perturbative computation at three-loop order~\cite{pelissetto01b}:
\begin{equation}
N_c(4-\epsilon)= 21.8 -23.4 \epsilon +7.09 \epsilon^2
+O(\epsilon^3)\ .
\label{Ncritique}
\end{equation} 
Unfortunately, this series is not well behaved since the coefficients
are not decreasing fast. It has been conjectured by Pelissetto et
al.~\cite{pelissetto01b} that $N_c(2)=2$. Using this conjecture, these
authors have reexpressed (\ref{Ncritique}) in the form:
\begin{equation}
N_c(4-\epsilon)=2+(2-\epsilon)(9.9-6.77\epsilon+0.16 \epsilon^2)
+O(\epsilon^3)\ .
\label{Ncritique_ameliore}
\end{equation} 

The coefficients of this expression are now rapidly decreasing so that
it can be used to estimate $N_c(d)$.  For $d=3$ it provides
$N_c(3)=5.3$ and leads to the conclusion that the transition is of
first order in the relevant Heisenberg and XY cases.

In agreement with this result, the perturbative approaches performed
at three loops, either in $4-\epsilon$ or directly in three
dimensions, lead to a first order phase transition for XY systems with
a $N_c(3)$ given respectively by $N_c(3)=3.91$ (Ref.\,\onlinecite{antonenko94}) and
$N_c(3)=3.39$ (Ref.\,\onlinecite{antonenko95}).  However, according to the authors,
these computations are not well converged. It is only recently that a
six-loop calculation has been performed
\cite{pelissetto00,pelissetto01a} directly in three dimensions which
is claimed to be converged in the Heisenberg and XY cases. Note that,
for values of $N$ between $N\simeq 5$ and $N\simeq 7$, the resummation
procedures do not lead to converged results, forbidding the authors to
compute $N_c(3)$ in this way.  For $N=2$ and $N=3$ a fixed point is
found. The exponents associated to the $N=2$ case are given in
Table~\ref{table_exp}. Note that $\gamma$ and $\nu$ compare reasonably
well with the experimental data of group 1. However, as we already
stressed, the existence of a fixed point implies $\eta>0$ ---
$\eta=0.08$ in Ref.\,\onlinecite{pelissetto01a} --- and is thus incompatible with the
negative value of $\eta$ found for the group~1. Moreover, the value
$\beta=0.31(2)$ found in Ref.\,\onlinecite{pelissetto01a} is far --- 4
standard deviations --- from the average experimental value
$\beta=0.237(6)$ for group~1 and also far --- 3.7 standard deviations
--- from that obtained from group~2, Ho and Dy: $\beta=0.388(7)$. It
is thus {\it incompatible} with the two sets of experimental values.
This point strongly suggests that the six-loop fixed point neither
describes the physics of materials belonging to group~2 that, in the
simplest hypothesis, should also undergo a first order phase
transition.

The preceding discussion does not rule out the existence of the fixed
point found in Ref.\,\onlinecite{pelissetto01a}. This just shows that,
if it exists, it must have a very small basin of attraction, and that
the initial conditions corresponding to the STA and helimagnets lie
out of it. In fact, as we argue in the following, this fixed point
probably does not exist at all so that we expect that all transitions
are of (possibly very weak) first order.

\section{The exact RG approaches}
There exists an alternative theoretical approach to the perturbative
RG calculations which explains well, qualitatively and to some extent
quantitatively, all the preceding facts. It relies on the Wilsonian RG
approach to critical phenomena, based on the concept of block spins
and scale dependent effective
theories\cite{kadanoff66,wilson74}. Although it has been originally
formulated in terms of hamiltonians, its most recent and successful
implementation involves the effective (average)
action\cite{wetterich93c,ellwanger94c,morris94a}. In the same way as
in the original Wilsonian approach, one constructs an effective
action, noted $\Gamma_k$, that only includes high-energy {\it
fluctuations} --- with momenta $q^2>k^2$ --- of the microscopic
system.  At the lattice scale $k=\Lambda=a^{-1}$, $\Gamma_k$
corresponds to the classical Hamiltonian $H$ since no fluctuation has
been taken into account. When the running scale $k$ is lowered,
$\Gamma_k$ includes more and more low-energy fluctuations. Finally,
when the running scale is lowered to $k=0$, {\it all} fluctuations
have been integrated out and one recovers the usual effective action
or Gibbs free energy $\Gamma$. To summarize one has:
\begin{equation}
\left\{
\begin{array}{ll}
\Gamma_{k=\Lambda}=H\ 
\\ 
\\
\Gamma_{k=0}=\Gamma\ .
\label{limitesgamma}
\end{array}
\right.
\end{equation}
Note also that the original hamiltonian depends on the original spins
while the effective action --- at $k=0$ --- is a function of the order
parameter. At an intermediate scale $k$, $\Gamma_k$ is a function of
an average order parameter at scale $k$ noted $\phi_k(q)$ --- or more
simply, $\phi(q)$ --- that only includes the fluctuations with momenta
$q^2>k^2$. Thus $\Gamma_k$ has the meaning of a free energy at scale
$k$.

At a generic intermediate scale $k$, $\Gamma_k$ is given as the
solution of an {\it exact} equation that governs its evolution with
the running scale~\cite{tetradis94}:
\begin{equation}
{\frac{\partial \Gamma_k[\phi]}{ \partial t}}=\frac1 2 \hbox{Tr}
 \left\{\left(\Gamma_k^{(2)}[\phi]+R_k\right)^{-1} 
 \frac {\partial R_k}{ \partial t}\right\}\; ,
\label{renorm}
\end{equation}
where $t=\ln \displaystyle {(k/\Lambda)}$ and the trace has to be
understood as a momentum integral as well as a summation over internal
indices. In Eq.(\ref{renorm}), $\Gamma_k^{(2)}[\phi]$ is the {\it
exact field-dependent} inverse propagator \ie the second functional
derivative of $\Gamma_k$.  The quantity $R_k$ is the infrared cut-off
which suppresses the propagation of modes with momenta $q^2<k^2$. A
convenient cut-off, that realizes the constraints
(\ref{limitesgamma}), is provided by~\cite{litim00}:
\begin{equation}
R_k(q^2)=Z (k^2-q^2)\theta(k^2-q^2)
\end{equation} 
where $Z$ is the $k$-dependent field renormalization.

Ideally, in order to relate the thermodynamical quantitites to the
microscopic ones, one should integrate the flow equation starting from
$k=\Lambda$ with the microscopic hamiltonian $H$ as an initial
condition and decrease $k$ down to zero. However, Eq.(\ref{renorm}) is
too complicated to be solved exactly and one has to perform
approximations. To render Eq.(\ref{renorm}) manageable, one truncates the
effective action $\Gamma_k[\phi]$ to deal with a finite number of
coupling constants. The most natural truncation, well suited to the
study of the long distance physics of a field theory, is to perform a
derivative expansion~\cite{tetradis94} of $\Gamma_k[\phi]$. This
consists in writing an {\em ansatz} for $\Gamma_k$ as a series in
powers of $\partial \phi$. The physical motivation for such an
expansion is that since the anomalous dimension $\eta$ is small, terms
with high numbers of derivatives should not drastically affect the
physics.

Actually, another truncation is performed. It consists in expanding
the potential, which involves all powers of the $O(N)\times O(2)$
invariants built out of $\vec{\phi}_1$ and $\vec{\phi}_2$, in powers
of the fields. This kind of approximation allows to transform the
functional equation (\ref{renorm}) into a set of ordinary coupled
differential equations for the coefficients of the expansion. It has
been shown during the last ten years that low order approximations in
the field expansion give very good results (see
Ref.\,\onlinecite{berges02} for a review and Ref.\,\onlinecite{bagnuls01}
for an exhaustive bibliography). The simplest such truncation
is~\cite{tissier00}:
\begin{equation}
\begin{split}
\Gamma_k = \int d^d x\Bigg\{&\frac Z 2 \hbox{Tr}\left(\partial
^{\;t}\!\Phi.  \partial\Phi\right)+\frac{\omega}{4} (\vec
\phi_1.\partial \vec\phi_2-\vec \phi_2.\partial \vec\phi_1)^2+
\\&+\frac{{\lambda}}{4}\left({\rho\over 2} - {\kappa}\right)^2 +
\frac{{\mu}}{4} \tau \Bigg\}\ .
\label{troncation}
\end{split}
\end{equation}

Let us first discuss the different quantities involved in this
expression. One recalls that $\Phi$ is the $N\times 2$ matrix
gathering the $N$-component vectors $\vec{\phi}_1$ and $\vec{\phi}_2$
(see Eq.(\ref{matriceparametreordre})). There are two  independent
$O(N)\times O(2)$ invariants given by $\rho={\hbox{Tr}}\,^{t}\Phi\Phi$ and $\tau=\frac{1}{
2}{\hbox{Tr}}\,(^{t}\Phi\Phi)^2-\frac{1}{4}({\hbox{Tr}}\
^{t}\Phi\Phi)^2$.  The set
$\left\{\kappa, \lambda, \mu,Z,\omega\right\}$ denotes the
scale-dependent coupling constants which parametrize the model at this
order of the truncation. The
first quantity in Eq.(\ref{troncation}) corresponds to the standard
kinetic term while the third and fourth correspond to the potential
part. Actually, apart from the second term --- called the current term
---, $\Gamma_k$ in Eq.(\ref{troncation}) looks very much like the usual
Landau-Ginzburg-Wilson action used to study perturbatively the
critical physics of the $O(N)\times O(2)$ model, up to trivial
reparametrizations. There is however a fundamental difference since we
do {\it not} use $\Gamma_k$ within a weak-coupling perturbative
approach. This allows the presence of the current term which
corresponds to a non-standard kinetic term. This term is irrelevant by
power counting around four dimensions since it is quartic in the
fields and quadratic in derivatives. However its presence is {\it
necessary} around two dimensions to recover the results of the
low-temperature approach of the nonlinear sigma (NL$\sigma$) model
since it contributes to the field renormalization of the Goldstone
modes.  Being not constrained by the usual power counting we include
this term in our ansatz. Note also that we have considered much richer
truncations than that given by Eq.(\ref{troncation}) by putting all the
terms up to $\Phi^{10}$ and by adding all terms with four fields and
two derivatives. This has allowed us to check the stability of our
results with respect to the field expansion. 

We do not provide the details of the computation. The general
technique is given in several publications and its implementation on
the specific $O(N)\times O(2)$ model will be given in a forthcoming
article\cite{tissier02b}.  The $\beta$ functions for the different
coupling constants entering in (\ref{troncation}) are given by:
\begin{widetext}
\begin{subequations}
\begin{align}
\begin{split}
\frac{d\kappa}{ dt}=&-(d-2+\eta)\kappa+4 v_d\bigg[\frac 1 2
 l_{01}^{d}(0, 0,\kappa \omega) + (N-2) l_{10}^d(0,0,0) + \frac3 2
 l_{10}^d(\kappa \lambda, 0, 0) + \left(1 + 2\frac{\mu}{
 \lambda}\right) l_{10}^d(\kappa \mu,
 0,0)+\\&+\frac{\omega}{\lambda}l_{0 1}^{2 + d}(0, 0, \kappa \omega)
 \bigg]
\label{flot_kappa_frustre}
\end{split}\displaybreak[0]\\  
%&\nonumber 
%\nonumber \\
\begin{split}
\frac{d\lambda }{dt}=&(d-4+2\eta)\lambda+ v_d \bigg[2\lambda^2 (N-2)
l_{20}^d( 0, 0, 0)+\lambda^2 l_{02}^d(0, 0, \kappa \omega) +9\lambda^2
l_{20}^d(\kappa\lambda, 0, 0) + 2 (\lambda + 2\mu)^2 l_{20}^d(
\kappa\mu, 0, 0) +\\&+4 \lambda\omega l_{02}^{2 + d}(0,
0,\kappa\omega)+ 4 \omega^2 l_{02}^{4 + d}(0, 0,\kappa
\omega)\bigg]
\label{flot_lambda_frustre}
\end{split}\displaybreak[0]\\
%&\nonumber \\
\begin{split}
\frac{d\mu }{dt}=&(d-4+2\eta)\mu -2 v_d\mu\bigg[-\frac{2}{\kappa}
l_{01}^d(0, 0, \kappa \omega) +\frac{3(2\lambda +
\mu)}{\kappa(\mu-\lambda)} l_{10}^d(\kappa\lambda,0,0) +\frac{8
\lambda + \mu}{ \kappa (\lambda - \mu)} l_{10}^d(\kappa \mu, 0,0)+\\& +
\mu l_{11}^d( \kappa \mu, 0,\kappa \omega) + \mu (N-2) l_{20}^d(0, 0,
0)\bigg]
\label{flot_mu_frustre}
\end{split}\displaybreak[0]\\
%&\nonumber 
%\\
\begin{split}
\eta=&-\frac{d\;\ln Z}{dt}=2\frac{v_d}{d\kappa} \bigg[(4-d)\kappa
 \omega l_{01}^d(0, 0,\kappa \omega) + 2\kappa^2 \omega^2 l_{02}^{2 +
 d}(0, 0, \kappa \omega) + 2 m_{02}^d(0, 0, \kappa \omega)- 4
 m_{11}^d( 0, 0, \kappa \omega)+\\&+2(-2 + d) \kappa \omega
 l_{10}^d(0, 0, 0)  +2 m_{20}^d(0, 0, \kappa \omega)+ 2 \kappa^2
 \lambda^2 m_{2, 2}^d(\kappa \lambda, 0, 0) +4 \kappa^2 \mu^2 m_{2,
 2}^d(\kappa \mu, 0, 0) + \\& + 4 \kappa \omega n_{02}^d(0, 0, \kappa
 \omega) -8 \kappa \omega n_{11}^d(0, 0, \kappa \omega) +4 \kappa
 \omega n_{20}^d(0, 0,\kappa \omega)\bigg]
\label{flot_Z_frustre}
\end{split}\displaybreak[0]\\
%&\nonumber 
%\\
\begin{split}
\frac{d\omega}{ dt}=&(d-2+ 2 \eta) \omega + \frac{4 v_d}{d \kappa^2}
\bigg[\kappa \omega\bigg\lbrace \frac{(4 -d)}{ 2} l_{01}^d(0, 0,
\kappa \omega) +\frac{(d-16 )}{2} l_{01}^d(\kappa \lambda, 0, \kappa
\omega) +\kappa \omega l_{02}^{2 + d}(0, 0, \kappa \omega) -\\&- 3
\kappa \omega l_{02}^{2 + d}(\kappa \lambda, 0, \kappa \omega)+(d-2)
l_{10}^d(0, 0, 0) - (d-8 ) l_{10}^d(\kappa \lambda, 0, 0)+8 \kappa
\lambda l_{11}^d(\kappa \lambda, 0, \kappa \omega) +2 \kappa \omega
l_{20}^{2 + d}(\kappa \mu, 0, 0) \\ &+2 \kappa \omega (N-2) l_{20}^{2
+ d}(0, 0, 0) \bigg\rbrace +m_{02}^d(0, 0, \kappa \omega) -
m_{02}^d(\kappa \lambda, 0, \kappa \omega) - 2 m_{11}^d(0, 0, \kappa
\omega) +2 m_{11}^d(\kappa \lambda, 0, \kappa \omega)+ 
\end{split}
\label{flot_omega_frustre}
\displaybreak[0]\\
\begin{split} \nonumber
& + m_{20}^d(0, 0,\kappa \omega)- m_{20}^d(\kappa \lambda, 0, \kappa
\omega)+\kappa^2 \lambda^2 m_{22}^d(\kappa \lambda, 0, 0) +2\kappa^2
\mu^2 m_{22}^d(\kappa \mu, 0, 0) + 2 \kappa \omega n_{02}^d(0, 0,
\kappa \omega)-\\& -4 \kappa \omega n_{02}^d(\kappa \lambda, 0, \kappa
\omega) - 4 \kappa \omega n_{11}^d(0, 0, \kappa \omega) +8 \kappa
\omega n_{11}^d(\kappa \lambda, 0, \kappa \omega) + 2 \kappa \omega
n_{20}^d(0, 0, \kappa \omega)  - 4\kappa \omega n_{20}^d(\kappa
\lambda, 0, \kappa \omega)\bigg]
\end{split}
\end{align}
\label{recursion}
\end{subequations}
\end{widetext}

In these equations appear the dimensionless functions
$l_{n1,n2}^d,m_{n1,n2}^d,n_{n1,n2}^d$, called threshold functions,
since they govern the decoupling of the massive modes entering in the
action (\ref{troncation}). They encode the nonperturbative content of
the flow equations (\ref{recursion}). They are complicated integrals
over momenta and are explicitly given in the Appendix.

\section{Results and physical discussion}
\subsection{Checks of the method}
We have first proceeded to all possible checks of our method by
comparing our results with all available data obtained within the
different pertubative approaches. Our method fulfills all these
checks.

1) Around $d=4$, we have checked that, in the limit of small coupling
constant, our equations degenerate in those obtained from the weak
coupling expansion at one loop\cite{jones76,bak76,garel76}:
\begin{equation}
\left\{
\begin{aligned}
{d\lambda\over dt} &= (d-4) \lambda +\frac{1}{16\pi^2}\left(4
\lambda\mu +4 \mu^2 +\lambda^2(N+4) \right)\\ {d\mu\over dt} &= (d-4)
\mu+\frac{1}{16\pi^2}\left(6 \lambda\mu + N \mu^2 \right).
\end{aligned}
\right.
\label{recursionglw}
\end{equation} 
This can be easily verified considering the asymptotic expressions of
the threshold functions that are given in the Appendix.

2) Also, around $d=2$, performing a low-temperature expansion ---
corresponding to a large $\kappa$ expansion --- of our equations and
making the change of variables:
\begin{equation}
\left\{
\begin{aligned}
\eta_1&=2\pi\kappa\\ \eta_2&=4\pi \kappa(1+\kappa\omega)\\
\end{aligned}
\right.
\end{equation}
one recovers the $\beta$-functions found in the framework of the
nonlinear $\sigma$ model at one loop\cite{azaria93}:
\begin{equation}
 \left\{
\begin{aligned} 
\beta_1&=- (d-2)\eta_1 +N-2-\frac{\eta_2}{2\eta_1}\\
\beta_2&=-(d-2)\eta_2
+\frac{N-2}{2}\left(\frac{\eta_2}{\eta_1}\right)^2 \ .
\ 
\end{aligned}
\right.
\label{recursionnls}
\end{equation}

This shows that our method allows to recover the perturbative results
both near $d=2$ and $d=4$. This is not a surprise since, as well
known, Eq.(\ref{renorm}) has a one-loop structure. However, for
frustrated magnets, and contrary to the $O(N)$ case, the matching with
the results of the NL$\sigma$ model is not trivial since it requires
to incorporate the non trivial current term which is irrelevant around
$d=4$, and thus absent in a LGW approach.

3) Our method also matches with the leading order results of the $1/N$
expansion of $\nu$ and $\eta$. In the $O(N)\times O(2)$ model these
exponents have also been computed at order $1/N^2$ in $d=3$. They are
given by \cite{pelissetto01b}:
\begin{equation}
 \left\{
\begin{aligned} 
\nu&=1-\frac{16}{\pi^2}\frac{1}{N}-\left(\frac{56}{\pi^2}
-\frac{640}{3\pi^4}\right) \frac{1}{N^2} +O(1/N^3)\\
\eta&=\frac{4}{\pi^2}\frac{1}{N}- \frac{64}{3\pi^4}\frac{1}{N^2}
+O(1/N^3)\ .
\end{aligned}
\right.
\label{nueta}
\end{equation}

We have computed $\nu$ and $\eta$ for a large range of values of $N$
to compare our results with those obtained within the $1/N$ expansion.
We meet an excellent agreement --- better than 1$\%$ --- for $\nu$
already for $N>10$ where the $1/N$ expansion is reliable.

4) For $N=6$, a Monte Carlo simulation has been performed on the STA
model\cite{loison00}. A second order phase transition has been found
without ambiguity with exponents given in Table
\ref{table_exp_crit_6_num}.  Our results compare very well with the
Monte Carlo data. Although the case $N=6$ does not correspond to any
physical system this constitutes a success of our approach from a
methodological point of view. Let us recall that, in this case, the
six loop weak coupling calculation is not converged.
\begin{table}[htbp]
\begin{tabular}{|l|l|l|l|l|l|}
\hline
&$\alpha$&$\beta$&$\gamma$&$\nu$&$\eta$\\
\hline
MC&-0.100(33)&0.359(14)&1.383(36)&0.700(11)&0.025(20)\\
\hline
P.W. &-0.121&0.372&1.377&0.707&0.053\\
\hline
\end{tabular}
\caption{Critical exponents for the $N=6$ STA system. The first row
corresponds to Monte Carlo data~{\cite{loison00}} and the second (P.W.)
to the present work.}
\label{table_exp_crit_6_num}
\end{table}

\begin{figure}[htbp] 
\centering
\includegraphics[width=0.85\linewidth,origin=tl]{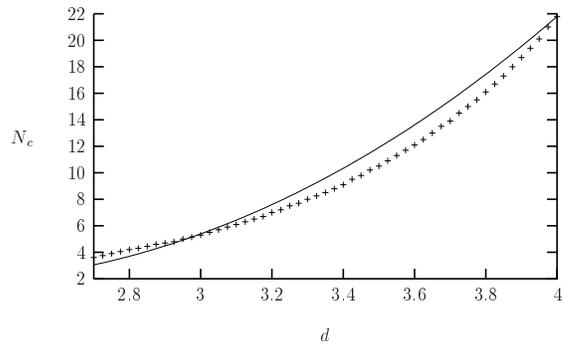}
\caption{The full line represents the curve $N_c(d)$ obtained by the
three-loop result improved by the constraint $N_c(2)=2$,
Eq.(\ref{Ncritique_ameliore}). The crosses represent our calculation.}
\label{nc_de_d}
\end{figure}

5) In agreement with the weak coupling perturbative results, we have
found by varying $N$ in a given dimension $d$ that, below a critical
value $N_c(d)$ of $N$, no fixed point exists~\cite{tissier00}(see
FIG. \ref{nc_de_d}). Our value of $N_c(d)$ agrees within ten percents
for {\it all} dimensions with the values obtained from
Eq.(\ref{Ncritique_ameliore}).  In $d=3$ we have found $N_c(3)=5.1$
which almost coincides with that obtained using
Eq.(\ref{Ncritique_ameliore}) that leads to $N_c(3)=5.3$.

All these checks show the consistency between our computation and most
of the previous theoretical approaches.  It is in contradiction with
the six loop calculation of Pelissetto {\it et al.}. Note that the
result $N_c(3)>3$ does not exclude {\it a priori} the existence of a
fixed point {\it not analytically} related --- in $N$ and
$\epsilon=4-d$ --- to those found above $N_c(d)$. This is, in fact,
the position advocated by Pelissetto {\it et
al.}~{\cite{pelissetto01b}}.  However, we have numerically searched
for this fixed point with our equations without success. This implies
three possibilities. The first one is that our method is not able ---
in principle --- to find this kind of fixed point.  Let us emphasize
that, although it is not possible to exclude this case, it is
improbable that a method which recovers all previous results in a
convincing way, misses such a fixed point. Another possibility is that
it has a so small basin of attraction that we have systematically
missed it in our numerical investigation of our equations. This would
mean that it probably does not play any role in the physics of
frustrated magnets.  The
  transitions should therefore be of first order for almost all real
  systems. It  would then be very difficult to
  explain the occurence of {\it generic weak first order} phase
  transitions.  Finally, there
remains the possiblity that the fixed point does not exist at all
which now appears as the most probable solution.

\subsection{Scaling without fixed and pseudo-fixed point}
The hypothesis of the absence of fixed point immediately raises the
theoretical challenge to explain the occurence of scaling in {\it
absence} of a fixed point. For Heisenberg spins, this question has
been addressed by Zumbach\cite{zumbach93} using a local potential
approximation (LPA) of the Polchinski equation\cite{polchinski84} and by the
present authors beyond LPA in the framework of the effective average
action method\cite{tissier00}.  It has been realized that, when $N$ is
lowered from $N>N_c(3)$ to $N<N_c(3)$, although the stable fixed point
disappears, there is no major change in the RG flow.

 This can be understood by considering  a domain  ${\cal E}$ of
  initial conditions of the flow corresponding to  all systems we are
  interested in ---  STA, STAR, $V_{N,2}$, real  materials, etc  ---  and studying  the  RG trajectories
 starting  in  ${\cal E}$.  One finds  that  there exists a
  domain ${\cal D}\subseteq{\cal E}$ such that all  trajectories
  emerging from ${\cal D}$ are attracted towards a {\it small}  domain  ${\cal R}$ in which the flow is very
  slow. Since  the flow is very slow in  ${\cal R}$ the RG time spent in this
region is long and, thus, the correlation lengths of the systems in
${\cal D}$ are very large. One therefore  partly recovers 
scaling for systems in  ${\cal D}$ (that aborts only for very small reduced
temperatures). Moreover, the smallness of ${\cal R}$ ensures the
existence of (pseudo-)universality. Consequently ${\cal R}$ mimics a true fixed
point.

This idea has been formalized through the concept of pseudo-fixed
point, corresponding to the point in ${\cal{R}}$ where the flow is the
slowest, the minimum of the flow\cite{zumbach93}.  At this point it
has been possible to compute (pseudo-)critical exponents
characterizing the pseudo-scaling (and pseudo-universality)
encountered in Heisenberg frustrated spin
systems\cite{zumbach93,tissier00}.

Within our approach, we confirm the existence, for values of $N$ just
below $N_c(3)$, of a minimum of the flow leading indeed to
pseudo-scaling and quasi-universality (see Ref.\,\onlinecite{tissier00}
and, for details, Ref.\,\onlinecite{tissier02b}). However, when $N$ is lowered, the minimum of
the flow is less and less pronounced and, for some value of $N$
between 2 and 3, it completely disappears.  Since (pseudo-)scaling is
observed in experiments and numerical simulations in XY systems this
means that the minimum of the flow does not constitute the ultimate
explanation of scaling in absence of a fixed point. At this stage, one
reaches the limits of the notion of minimum of the flow as the
quantity playing the role of (pseudo-)fixed point.  First, it darkens
the important fact that the notion relevant to scaling is not the
existence of a minimum of the flow but that of a whole region
${\cal{R}}$   in which the flow is slow, \ie the $\beta$ functions
are small. Put it differently, the existence of a minimum of the flow
does not guarantee that the flow is slow, {\ie}that the correlation
length is large compared with the lattice spacing. Reciprocally, it
can happen that the RG flow is slow, the correlation length being
large, and that scaling occurs even in absence of a minimum.  Second,
reducing $\cal{R}$ to a point, one rules out the possibility to test
the violation of universality.  Clearly, the degree of universality is
related to the size of $\cal{R}$. The smaller $\cal{R}$ the more
universal is the behavior.

 Qualitatively --- thanks to continuity arguments --- one expects that,
for $N$ close to $N_c(3)$,  all systems in $\cal E$ exhibit
  pseudo-scaling
--- ${\cal D}={\cal E}$ ---  and ${\cal{R}}$ is
almost pointlike so that the transitions are extremely weakly of first
order and universality almost holds. When $N$ is  sufficiently
decreased,   two phenomena occur. First, 
${\cal{D}}$ gets smaller  than ${\cal E}$ and thus the transitions
become of strong first order for all systems that belong to  ${\cal
  E}$ but not to  ${\cal D}$.
Second,  ${\cal{R}}$ gets wider and thus  a whole spectrum of exponents is
observed and  pseudo-universality breaks down. Note that these two phenomena
are not obligatorily simultaneous so that, for intermediate values of
$N$, scaling still holds while pseudo-universality is already significantly violated.
 These two phenomena  are  observed numerically in the  Heisenberg and
 XY systems: STA, STAR and $V_{N,2}$ models. For $N=3$, all these models  display scaling but with $\beta$
exponents that are almost  incompatible [$\beta_{STA}$=0.285(11),\cite{mailhot94}  
$\beta_{STAR}$=0.221(9) and $\beta_{V_{3,2}}$=0.193(4) (see
  Ref.\,\onlinecite{loison00b})] which means that one probably starts
leaving  the pseudo-universal regime.  For $N=2$, scaling is
only observed for STA --- $\beta=0.24(2)$ here --- while the
transitions for STAR and $V_{2,2}$ are found to be of first order.

From a theoretical point of view, the Heisenberg case has been already
treated in Ref.\,\onlinecite{tissier00}. We now concentrate on the XY
case.
 
\subsection{Our results}
In practice, we numerically integrate the flow equations
(\ref{recursion}) and compute physical quantities like correlation
length and magnetization as a function of the reduced temperature
$t_r=(T-T_c)/T_c$. Since one expects the behavior of frustrated
magnets to be nonuniversal, one should study each system independently
of the others. Thus, ideally, one should consider as initial
conditions of the RG flow all the microscopic parameters
characterizing a {\it specific} lattice system.  This program would
require to identify and to deal with an infinite number of coupling
constants. This remains a theoretical challenge. Rather than doing
this, we have chosen to address the question of scaling in absence of
a fixed point independently of a given microscopic system, using for
$\Gamma_k$ a finite {\it ansatz} similar to, but richer than,
Eq.({\ref{troncation}).

Since our truncations forbid us to relate precisely the
thermodynamical quantities to the microscopic couplings, we have
chosen, as initial conditions, the simplest temperature dependence of
the parameters at the scale $k=\Lambda$. This consists in fixing all
coupling constants to temperature-independent values and in taking the
usual {\it ansatz} for the quadratic term of $\Gamma_k$:
\begin{equation}
\kappa_{k=\Lambda}=a+b\,T
\end{equation}
where $a$ and $b$ are parameters that we have varied to test the
robustness of our conclusions. For each temperature we have integrated
the flow equations and deduced the $t_r$-dependence of the physical
quantities around the critical temperature.

Let us now review the main results obtained by the integration of the
RG flow.

1) The integration of the flow equations leads {\it generically} to
good power laws in reduced temperature for the magnetization, the
correlation length (see FIG. \ref{log_m_xi_neq2}) and the
susceptibility. We are thus able to extract pseudo-critical exponents
varying typically, for $\beta$ between 0.25 and 0.38 and for $\nu$
between 0.47 and 0.58.  This phenonomenon holds for a wide domain
${\cal D}$ in coupling constant space.

\begin{figure}[tbp] 
\centering
\makebox[\linewidth]{
\label{log_aimantation_neq2}
\includegraphics[width=.95\linewidth,origin=tl]{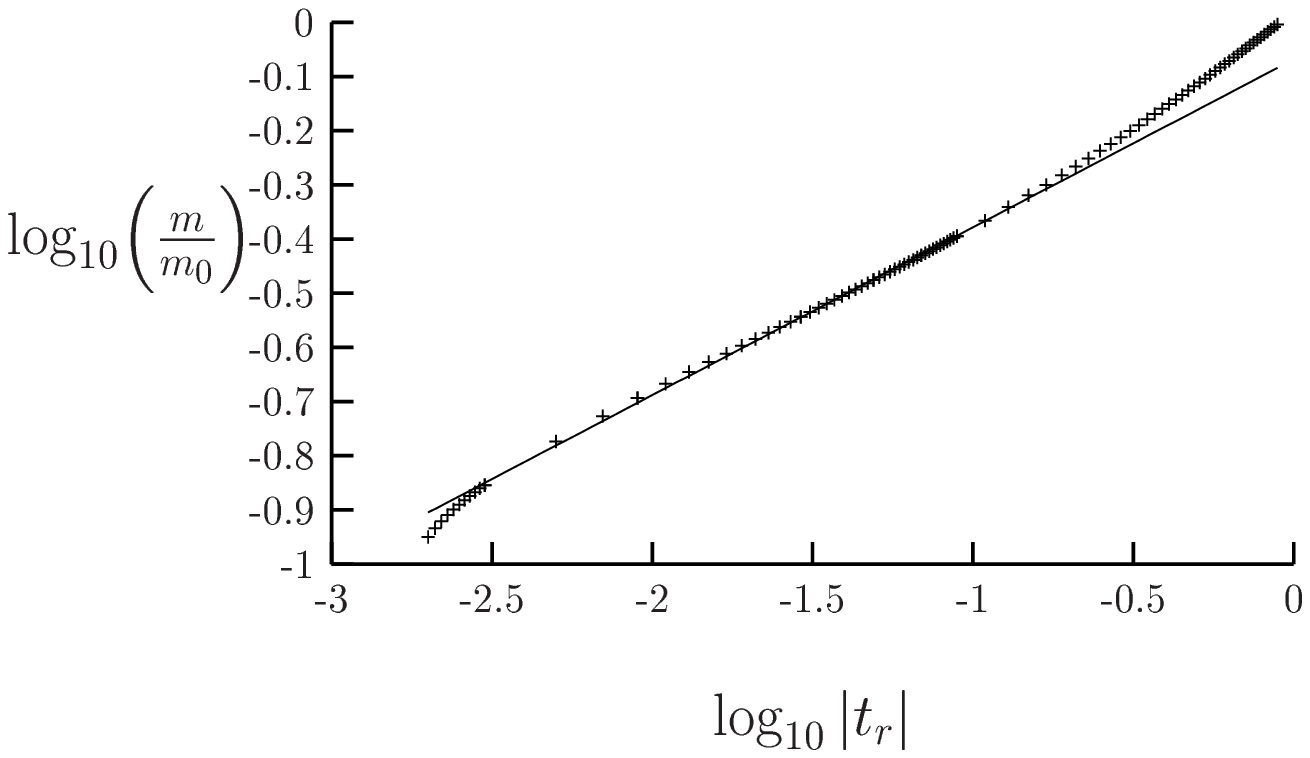}}
\makebox[\linewidth]{
\label{log_correlation_neq2}
\includegraphics[width=.95\linewidth,origin=tl]{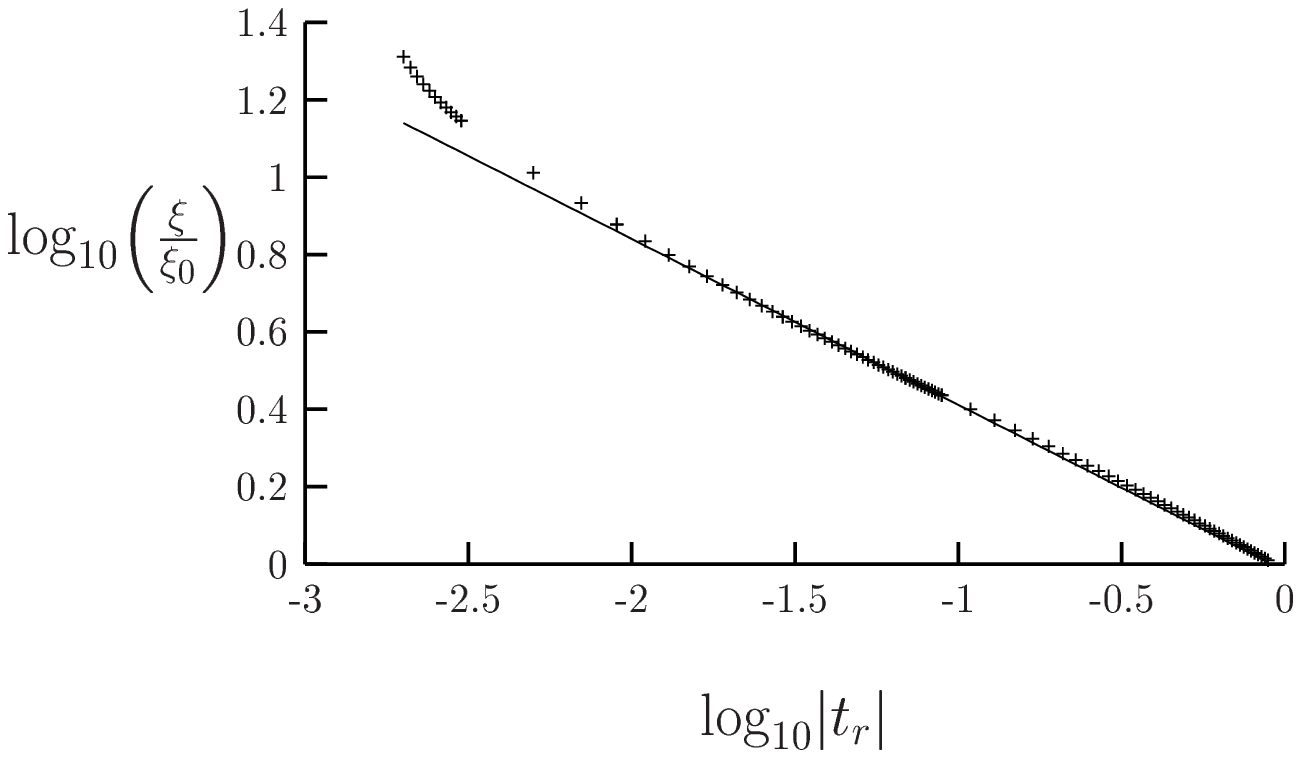}% 
}
\caption{Log-log plots of the magnetization $m$ and of the correlation
length $\xi$ for $N=2$ as functions of the reduced temperature
$t_r$ for parameters corresponding to materials of group 1. The
straight  lines correspond to the best power law fit of the
data. The power-law behavior observed breaks down for small $t_r$.}
\label{log_m_xi_neq2}
\end{figure}

2) We easily find initial conditions leading to pseudo-exponents close
to those of group 2: $\beta=0.38$, $\nu=0.58$ and $\gamma=1.13$ (see
Ho and Dy in Table~\ref{table_exp} for comparison). We have checked
that the previous result is quite stable to both variations of the
microscopic parameters and to a change of the $T$-dependence of the
microscopic coupling constants. This is in agreement with the
stability of $\beta$ in group~2. In the region of parameters leading
to this behavior, correlation lengths as large as 5000 lattice
spacings are found.

3)   We also easily  find initial conditions leading to $\beta \simeq 0.25$,
corresponding to group~1. The power laws then hold on smaller range of
temperature and the critical exponents are more sensitive to the
determination of $T_c$ and to the initial conditions, in agreement
with point {\it ii)} of Section III-A. For such values of $\beta$, we find a range
of values of $\nu$ --- $0.47\leq\nu\leq0.49$ --- which is somewhat below
the value found for CsMnBr$_3$ (see Table~\ref{table_exp}). Also,
the corresponding $\eta$ deduced from the scaling relation
$\eta=2\beta/\nu-1$ is always positive and, at best, zero. Finally, it is very
interesting to notice that when we find $\beta$ of order 0.25 (group
1) we also find correlation lengths at the transition of the order of
a few hundreds lattice spacings which coincide rather well with the
first size where a direct evidence of a first order transition has
been seen in Monte Carlo RG simulations (lattice sizes around
100)~\cite{itakura03}.
 
4) For $\beta\simeq 0.3$ we find critical exponents in good agreement
with those obtained by the six-loop calculation
of Ref.\,\onlinecite{pelissetto01a} (see Table~\ref{table_exp}). For
instance, for $\beta=0.33$ we find $\nu\simeq 0.56$ and $\gamma\simeq
1.07$.

\section{Conclusion}
On the basis of their specific symmetry breaking scheme it has been
proposed~\cite{garel76,yosefin85,kawamura88} that the critical physics
of XY frustrated systems in three dimensions could undergo a second
order phase transition characterized by critical exponents associated
with a new universality class.  We have given convincing arguments
that rather favor the occurence of --- generically weak --- first
order transitions for all XY frustrated magnets with a spreading of
(pseudo-)critical exponents. This is supported by experimental and
numerical results that do not agree with a second order
behavior. Moreover we have shown, using a nonperturbative approach,
that this generic but nonuniversal scaling finds a natural explanation
in terms of slowness and ``geometry'' of the flow. Our approach
appears to explain the main puzzling features of XY frustrated
magnets.

We now propose several tests to confirm our approach. On the
experimental side, more accurate determinations of critical exponents
could lead to a definitive answer on the nature of the transition, at
least if no drastically new physics emerges (as it could be the case
for Helimagnets). In particular, it is important to check that $\eta$
is negative for CsMnBr$_3$, and therefore to refine the determination
of $\nu$. It would also be of utmost interest to have more precise
determinations of $\alpha$ in CsMnBr$_3$, CsNiCl$_3$ and CsMnI$_3$
which are, up to now, only marginally compatible. In case they are
different, this would corroborate the lack of universality that we
predict. The existence of a continuous spectrum of critical exponents
could be directly tested by simulating, for example, a family of
models, extrapolating continuously from STA to STAR.

On the theoretical side, it would be of interest to push the
derivative expansion to refine the value of $\eta$ for group~1, which
is, as usual, overestimated~\cite{tetradis94}. This would allow to
reproduce its observed negative value.  Moreover, it would be
interesting to clarify the discrepancy between the nonperturbative
approach and the six loop result. The ability of our approach to
reproduce exactly the whole set of exponents found by Pelissetto {\it
et al.} suggests that their fixed point does not correspond to a true
fixed point but, in fact, to a region where the flow is very slow.
One can thus question the convergence of the perturbative result which
is not Borel summable. A detailed analysis of this problem of
convergence could reveal that the real fixed point found in the
perturbative approach is, actually, a complex one.

Finally, the major characteristics of XY-frustrated magnets {\it i.e.}
the existence of scaling laws with continuously varying exponents are
probably encountered in other physical contexts, generically systems
with a critical value $N_c$ of the number of components of the order
parameter, separating a true second order behavior and a na\"{\i}vely
first order one.  \acknowledgements{We thank P. Schwemling for useful
discussions. Laboratoire de Physique Th\'eorique et Hautes Energies,
Universit\'e Paris VI Pierre et Marie Curie --- Paris VI Denis Diderot
--- is a laboratoire associ\'e au CNRS UMR 7589. Laboratoire de
Physique Th\'eorique et Mod\`eles Statistiques, Universit\'e Paris
Sud, is a laboratoire associ\'e au CNRS UMR 8626.

\appendix
\section{Threshold functions}
\label{annexe_threshold}

We discuss in this appendix the different threshold functions
$l_{n_1,n_2}^d$, $m_{n_1,n_2}^d$ and $n_{n_1,n_2}^d$ appearing in the
flow equations (\ref{recursion}).

The threshold functions are defined as:
\begin{widetext}
\begin{subequations}
\begin{align}
l_{n_1,n_2}^d(w_1,w_2,a)&=-\frac12\int_0^\infty dy\; y^{d/2-1}
\tilde{\partial}_t\left\{\frac{1}{(P_1(y)+w_1)^{n_1}
(P_2(y)+w_2)^{n_2}}\right\} ,\label{annexe_def_l}\\ \displaybreak[0]
m_{n_1,n_2}^d(w_1,w_2,a)&=-\frac12\int_0^\infty dy\; y^{d/2-1}
\tilde{\partial}_t\left\{\frac{y(\partial_y
P_1(y))^2}{(P_1(y)+w_1)^{n_1} (P_2(y)+w_2)^{n_2}}\right\}
\label{annexe_def_m},\\\displaybreak[0]
n_{n_1,n_2}^d(w_1,w_2,a)&=-\frac12\int_0^\infty dy\; y^{d/2-1}
\tilde{\partial}_t\left\{\frac{y\partial_y P_1(y)}{(P_1(y)+w_1)^{n_1}
(P_2(y)+w_2)^{n_2}}\right\}\label{annexe_def_n}\ 
\end{align}
\label{annexe_def_thres}
\end{subequations}
\end{widetext}
with:
\begin{align}
P_1(y)&=y(1+r(y)+a)\\ P_2(y)&=y(1+r(y))\ .
\end{align}
In all the previous expressions $y$ is a dimensionless quantity:
$y=q^2/k^2$ where $q$ is the momentum variable over which the integral
in Eq.(\ref{renorm}) is performed. As for $r(y)$, it corresponds to the
dimensionless renormalized infrared cut-off:
\begin{equation}
r(y)=\frac{R_k(q^2)}{Z q^2}=\frac{R_k(y k^2)}{Z k^2 y}\ .
\end{equation}

In Eqs.\,\ref{annexe_def_l}--\ref{annexe_def_n}, $\tilde\partial_t$ means
that only the $t$-dependence of the function $R_k$ is to be considered
and not that of the coupling constants. Therefore one has:
\begin{align}
\tilde{\partial}_t P_i(y)&=\frac{\partial R_k}{\partial t}
\frac{\partial}{\partial R_k} P_i(y)\\
&=-y(\eta r(y)+2y r'(y)).
\end{align}
Now the threshold functions can be expressed as explicit integrals if
we compute the operation of $\tilde \partial_t$. To this end, it is
interesting to notice the equality: $\partial_y \tilde{\partial}_t
P_i(y) = \tilde{\partial}_t \partial_y P_i(y)$, so that:
\begin{equation}
\tilde{\partial}_t\partial_y r=-\eta(r+yr')-2y(2r'+yr'')
\end{equation}
We then get:
\begin{widetext}
\begin{gather}
\begin{split}
l_{n_1,n_2}^d(w_1,w_2,a)=-\frac12\int_0^\infty dy\; y^{d/2}
\;&\frac{\eta r +2yr'}{(P_1(y)+w_1)^{n_1}
(P_2(y)+w_2)^{n_2}}\left(\frac{n_1}{P_1(y)+w_1}+\frac{n_2}{P_2(y)+w_2}
\right)
\end{split}\\ \displaybreak[0]
\begin{split}
n_{n_1,n_2}^d(w_1,w_2,a)=-&\frac12\int_0^\infty dy\; y^{d/2} \frac{1}
{(P_1(y)+w_1)^{n_1}(P_2(y)+w_2)^{n_2}} \Bigg\{y(1+a+r+yr')(\eta r+2y
r')\times\\&\left(\frac{n_1}{P_1(y)+w_1} +\frac{n_2}{P_2(y)+w_2}
\right)-\eta(r+yr')-2y(2r'+yr'')\Bigg\}
\end{split}\\ \displaybreak[0]
\begin{split}
m_{n_1,n_2}^d(w_1,w_2,a)=-&\frac12\int_0^\infty dy\; y^{d/2}
\frac{1+a+r+yr'}
{(P_1(y)+w_1)^{n_1}(P_2(y)+w_2)^{n_2}}\Bigg\{y(1+a+r+yr')(\eta r+2y
r')\times\\&\Bigg(\frac{n_1}{P_1(y)+w_1}+\frac{n_2}{P_2(y)+w_2}
\Bigg)-2\eta(r+yr')-4y(2r'+yr'') \Bigg\}\ .
\end{split}
\end{gather}
\end{widetext}

\end{document}